\def\year{2022}\relax
\documentclass[journal]{IEEEtai} 
\usepackage[hyphens]{url}  
\usepackage{graphicx} 
\usepackage{amsmath}
\usepackage{booktabs}
\usepackage{color}

\usepackage{multirow}
\graphicspath{{images/}} 

\title{Fine-grained Early Frequency Attention for Deep Speaker Representation Learning} 

\author{
Amirhossein Hajavi\textsuperscript{}\thanks{The authors would like to thank IMRSV Data Labs for their support of this work. The authors would also like to acknowledge the Natural Sciences and Engineering Research Council of Canada (NSERC) for supporting this research (grant number: CRDPJ 533919-18). }, Ali Etemad \textsuperscript{} \\ 

\textsuperscript{}Department of ECE and Ingenuity Labs Research Institute\\Queen's University, Kingston, Canada\\

\{a.hajavi, ali.etemad\}@queensu.ca
}

\begin{document}

\maketitle

\begin{abstract}
Deep learning techniques have considerably improved speech processing in recent years. Speaker representations extracted by deep learning models are being used in a wide range of tasks such as speaker recognition and speech emotion recognition. Attention mechanisms have started to play an important role in improving deep learning models in the field of speech processing. Nonetheless, despite the fact that important speaker-related information can be embedded in individual frequency-bins of the input spectral representations, current attention models are unable to attend to fine-grained information items in spectral representations. In this paper we propose Fine-grained Early Frequency Attention (FEFA) for speaker representation learning. Our model is a simple and lightweight model that can be integrated into various CNN pipelines and is capable of focusing on information items as small as frequency-bins. We evaluate the proposed model on three tasks of speaker recognition, speech emotion recognition, and spoken digit recognition. We use Three widely used public datasets, namely VoxCeleb, IEMOCAP, and Free Spoken Digit Dataset for our experiments. We attach FEFA to several prominent deep learning models and evaluate its impact on the final performance. We also compare our work with other related works in the area. Our experiments show that by adding FEFA to different CNN architectures, performance is consistently improved by substantial margins, and the models equipped with FEFA outperform all the other attentive models. We also test our model against different levels of added noise showing improvements in robustness and less sensitivity compared to the backbone networks.
\end{abstract}



\section{Introduction}

Audio recordings of human speech often contain specific features from which a number of characteristics can be determined regarding the speaker. Speaker identity and emotional state are two of such characteristics that have been the subject of a number studies in recent years \cite{hansen_speaker_2015, AKCAY2020emotion}. Analyzing speaker-related information for applications such as \textit{Speaker Recognition (SR)} and \textit{Speech Emotion Recognition (SER)} often requires the use of spectral representations of audio signals. These frequency features undergo different processes inside automated systems to generate a representation of the utterance, namely the \textit{speaker representation}. The speaker representations ideally contain all the information required for performing SR or SER. In classical solutions, i-Vector systems \cite{kenny2007joint} generated these speaker representations under the name of identity vectors.

More recently, with advancements in Deep Neural Networks (DNN), speaker representation learning through DNNs has gained considerable attention. As a result, a significant number of DNNs have been explored for SR \cite{hajavi2019deep, chung_voxceleb} and SER \cite{gideon2019improving, wang2020speech}, respectively. The quality of speaker representations generated by DNNs far surpasses the ones generated by i-Vector systems \cite{nagrani_VoxCeleb, ghosh2016representation}, making DNNs the predominant solutions for speaker-related tasks.

Attention mechanisms have been an integral part of the recent advancements in deep learning \cite{chaudhari2021attentive, galassi2020attention}. 
To describe the general process of Attention mechanisms, they take information items as input and assign learnable weights to each item \cite{chaudhari2021attentive}. This process helps deep learning models to select the most valuable items to use further down in the pipeline, for instance for classification or regression tasks. Some examples of information items commonly used in attention mechanisms in SR and SER are the embeddings obtained from Convolutional Neural Networks (CNN) \cite{bian2019self, bhattacharya2017deep}, Recurrent Neural Network (RNN) \cite{wang2020speech, zhang_seq2seq_2019}, or Time-Delay Neural Network (TDNN) \cite{okabe_attentive_2018, zeinali_how_2019}. 

An attention mechanism calculates the aforementioned weights during the training of the model using back-propagation by receiving gradients from the layer that follows it in the DNN \cite{chaudhari2021attentive}. In the context of generating speaker representations from audio signals, the information items used by the majority of attention mechanisms \cite{bian2019self,  you2019deep, safari2019self, kye2021supervised, jung2021attention_speaker, india2021attention_speaker, mingote2021attention_speaker, wang2020speech, zhang_seq2seq_2019, zeinali_how_2019} are the embeddings generated by encoding the input via a CNN module while the gradients are usually derived from the final layer of the model. Using the CNN embeddings of the utterances as information items, makes it so that every information item corresponds to a specific region of the input.

The most commonly used input for speaker representation learning is the spectral representations of an utterance, in which each individual input value is a frequency-bin. However, despite the fact that specific speaker-related information may be contained in individual frequency-bins, existing attention mechanisms have not yet treated these individual bins as information items. Instead, these mechanisms have so far used particular \textit{regions} \cite{bian2019self, bhattacharya2017deep, wang2020speech, zhang_seq2seq_2019, okabe_attentive_2018, zeinali_how_2019, zhu_self-attentive_2018} of the input spectral representations as information items, thus suffering from low granularity. 

Improving the granularity of CNN embeddings used as information items results in a drastic increase in the dimensionality of the embeddings themselves and in turn leads to very large attention models. Such large attention models are generally hard to train \cite{chaudhari2021attentive, galassi2020attention}. Hence, despite the abundance of studies investigating various attention models for speaker representation learning \cite{bhattacharya2017deep, bian2019self, you2019deep, safari2019self, kye2021supervised, jung2021attention_speaker, india2021attention_speaker, mingote2021attention_speaker}, very limited number of works have aimed to use more fine-grained attention models for SR and SER. 

In this paper, we address the challenge of improving granularity of attention mechanisms for speaker representation learning by introducing the Fine-grained Early Frequency Attention (FEFA). This mechanism enables deep learning models to focus on individual frequency-bins of spectrogram representations without the drawbacks of having very complex attention models that typically involve many parameters. The aim of this model is to attend to each frequency-bin in the spectrogram representation in order to boost the contribution of salient bins. This mechanism also helps reduce the importance of bins with no useful information which in turn leads to more accurate representations, and thus leading to more robustness to the presence of noise.
We study the impact of the proposed attention mechanism on the performance of different backbone models for two widely used datasets, VoxCeleb \cite{chung_voxceleb} and IEMOCAP \cite{busso2008iemocap}, for SR and SER respectively. Additionally we experiment with the Free Spoken Digit Dataset (FSDD) \cite{amnist} dataset as a simple additional dataset in order to fully analyze the different aspects of our work. The experimental results show that deploying FEFA in different models improves the performance of all the benchmark networks substantially while being less impacted by added noise.

Our contributions in this paper are as follows:
\begin{itemize}
    \item We introduce a novel attention mechanism, FEFA, for speaker representation learning. Our model is a simple, yet effective module that can be easily integrated into existing DNN pipelines used for audio representation learning to boost their performance without requiring any modification to the DNN architecture.
    
    \item We evaluate our method on two different speaker-related problem domains, namely SR and SER, using two large and widely used datasets, demonstrating considerable performance gains for both tasks. Our analysis shows that FEFA improves the quality of the obtained embeddings and boosts performance with little added complexity.
    
    \item By testing our model against different levels of synthetic noise, we show an improvement in robustness compared to other models. 
\end{itemize}

This paper is an extension of our work on `Early Frequency Attention', presented at IJCNN 2022. The added contributions of this paper compared to the conference version can be summarized as follows: 
(1) We enhance the mechanism of the attention module by adding fully connected layers in addition to locally connected layers. 
(2) We add additional analysis and experiments on 2 other tasks of spoken digit recognition and speech emotion recognition in order to show the functionality and generalizability of our model. 
(3) We analyze the complexity of the proposed model and compare it to other forms of the attention mechanisms and show that despite using fewer number of parameters, our approach outperforms other solutions.

The rest of this paper is organized as follows. First, we discuss the related work in the area of speech representation learning with a focus on speaker-related tasks and explore particular approaches that have used attention mechanisms for this purpose. Next, we present our proposed method. Following, we discuss the experiments along with implementation details. Next, we provide the results and analysis of our experiments. And finally, we summarize and conclude the paper.

\section{Related Work}
\subsection{Speaker Representation Learning}
Extracting speaker representations or utterance embeddings, with the aim of acquiring speaker related information, has been extensively studied over the years. Classical techniques such as Gaussian Mixture Models \cite{burget2007analysis}, Hidden Markov Models \cite{campbell2003svm}, and Universal Background Models \cite{omar2010training}, have been used to obtain effective utterance representations that contain important information regarding the speaker of utterances. Comprehensive reviews of prior works that have used such conventional methods, particularly for SR and SER, can be found in \cite{hansen_speaker_2015,el2011survey}. 

Along with the classical techniques, some attempts were made to utilize artificial neural networks (ANN) for SR and SER. In some of the earlier works in this area, utterance embeddings extracted from audio signals using ANNs were fed to conventional classifiers to perform SR \cite{farrell1994speaker} and SER \cite{nicholson2000emotion}. The ANN-based solutions were able to achieve better performances compared to previous classical techniques. However, with the introduction of methods such as i-Vectors \cite{kenny2007joint} and the computational requirements of ANNs, further studies did not consider ANNs as a viable option.

Recent advancements in DNNs has led to a renewed interest in the use of such methods for learning effective representations of utterances for SR and SER. Most recent works on extracting deep utterance embeddings for these tasks have explored the impacts of different deep learning architectures on the quality of these representations \cite{bhattacharya2017deep, bian2019self, you2019deep, safari2019self, kye2021supervised, jung2021attention_speaker, india2021attention_speaker, mingote2021attention_speaker}. Most prominent works use CNN architectures such as ResNets for embedding speech representations towards performing SR \cite{xie2019utterance, bian2019self, hajavi2019deep} and using a combination of CNNs and RNNs such as long short-term memory (LSTM) for SER \cite{xie2019speech, latif2019direct, wang2020speech}. 

\begin{table*}[t]
\label{tab:litreview}
\centering
\caption{A Review of the current literature on the use of attention mechanisms for SR and SER. MFCC: Mel Frequency Cepstral Coefficients; PLP: Perceptual Linear Predictive; LFB: Log Filter Bank; FBN: Phonetic Bottleneck; Spect: Spectrogram; LLD: Low-Level Descriptors; eGeMaps: The Geneva Minimalistic Acoustic Parameter Set. }
\begin{tabular}{llclllllcc}\hline
Study                                               & Attention  & SR/SER & Model  & Dataset        & Wild      & Input           & Loss         & Early & Fine-grained \\ \hline \\[-.35cm]
Tarantino \cite{tarantino2019self}                  & Self       & SER    & Transformer    & IEMOCAP        & No      & eGeMaps           & Softmax      & No    & No           \\
Xie et al. \cite{xie2019speech}                     & Self       & SER    & LSTM   & Multiple       & No      & Spect-93          & Softmax      & No    & No           \\
Ramet et al. \cite{ramet2018context}                & Self       & SER    & RNN    & IEMOCAP        & No      & MFCC-12 + LLD-32  & Softmax      & No    & No           \\
Mirsamadi et al. \cite{mirsamadi2017automatic}      & Local      & SER    & RNN    & IEMOCAP        & No      & MFCC-12 + LLD-32  & Softmax      & No    & No           \\
Zhao et al. \cite{zhao2020end}                      & Temporal   & SER    & ResNet & VideoEmotion   & No      & MFCC-40 + Video   & Softmax      & No    & No           \\ \hline
Okabe et al. \cite{okabe_attentive_2018}            & Self       & SR     & TDNN   & VoxCeleb       & Yes      & MFCC-20           & Softmax      & No    & No           \\
Zhu et al. \cite{zhu_self-attentive_2018}           & Self       & SR     & TDNN   & SRE 2016       & Yes      & MFCC-23           & Softmax      & No    & No           \\
Zeinali et al. \cite{zeinali_how_2019}              & Self       & SR     & TDNN   & SRE 2018       & Yes      & MFCC-23           & Softmax      & No    & No           \\
Bian et al. \cite{bian2019self}                     & Self       & SR     & ResNet & VoxCeleb       & Yes      & FB-64             & Triplet      & No    & No           \\
India et al. \cite{safari2019self}                  & Multi-head & SR     & CNN    & VoxCeleb       & Yes      & Spect-128         & Softmax      & No    & No           \\
India et al. \cite{india2021attention_speaker}      & Multi-head & SR     & CNN    & VoxCeleb       & Yes      & Spect-80          & Softmax      & No    & No           \\
Wang et al. \cite{wang2020attention_speaker}        & Multihead  & SR     & ResNet & VoxCeleb       & Yes      & MFCC-40           & CosAMS       & No    & No           \\
Mingote \cite{mingote2021attention_speaker}         & Multi-head & SR     & ResNet & DeepMine       & Yes      & MFCC-20           & Softmax      & No    & No           \\
You et al. \cite{you2019deep}                       & Gated      & SR     & CNN    & SRE 2018       & Yes      & PLP-39            & Softmax      & No    & No           \\
Zhang et al. \cite{zhang_seq2seq_2019}              & Seq2Seq    & SR     & CRNN   & Tencent WU     & No       & MFB-128           & Binary       & No    & No           \\
Shi et al. \cite{shi2020attention_speaker}          & Multiple   & SR     & CRNN   & SwitchBoard    & No       & MFCC-20           & Softmax      & No    & No           \\
Chowdhury et al. \cite{chowd2018attention_speaker}  & Soft       & SR     & LSTM   & Google WU      & No       & MFCC-40           & Triplet      & No    & No           \\
Zhou et al. \cite{zhou2019attention_speaker}        & Phonetic   & SR     & ResNet & VoxCeleb       & Yes      & LFB-80 + PBN      & Softmax      & No    & No           \\
Kye et al. \cite{kye2021supervised}                 & Supervised & SR     & ResNet & VoxCeleb       & Yes      & MFB-40            & PL + Softmax & No    & No           \\
Jung et al. \cite{jung2021attention_speaker}        & Graph      & SR     & ResNet & VoxCeleb       & Yes      & MFCC-40           & Costume      & No    & No           \\
Yadav et al. \cite{yadav2020frequency}              & CBAM       & SR     & ResNet & VoxCeleb       & Yes      & Spect-256         & ArcSoftmax   & Yes   & No           \\

\hline
\end{tabular}
\end{table*}

\subsection{Attention Mechanisms for Speaker Representation Learning}
Attention mechanisms have been used to improve the performance of deep learning models \cite{devlin2019bert, xu2015show, bahdanau2015neural}. Subsequently, a large number of studies using attention mechanisms for SR and SER have shown substantial improvements compared to baseline models. These improvements have been achieved by simply utilizing such mechanisms to focus on features extracted from utterances by deep learning models with various architectures including CNN \cite{bian2019self, bhattacharya2017deep, you2019deep, safari2019self, zhao2020end, mingote2021attention_speaker}, RNN \cite{zhang_seq2seq_2019, wang2020speech, tarantino2019self, shi2020attention_speaker, chowd2018attention_speaker}, and time-delay neural networks (TDNN) \cite{okabe_attentive_2018, zeinali_how_2019, zhu_self-attentive_2018}. Through the following paragraphs we briefly describe the attention mechanisms used in these studies.

The model proposed in \cite{bhattacharya2017deep} utilized a self-attention mechanism to focus on features obtained from a CNN model inspired by VGGNet \cite{simonyan2014very} to generate utterance embeddings with more useful information for performing SR. The study done in \cite{bian2019self} used CNN layers inside a self-attention mechanism to attend to features extracted from the utterance using a deep learning model based on the ResNet architecture \cite{he_deep_2016}. The work done in \cite{you2019deep} proposed a gated attention mechanism that focused on features extracted from the utterance by a modified version of CNN, namely gated-CNN. In all of these studies, the use of attention mechanisms has enabled the models to focus on particular features extracted by the DNN, resulting in enhanced embeddings in comparison to the non-attentive solutions for SR.

Alternatively, the proposed models in \cite{zhang_seq2seq_2019, wang2020speech, tarantino2019self}, utilized attention mechanisms to focus on differences between two sets of extracted utterance embeddings. These embeddings were extracted from the reference and test utterances using RNNs. In the common approach taken in these studies, attention models were added to the deep learning pipelines in the backend of the speaker verification system where the final similarity score is calculated. These studies showed that utilizing attention models in this way helps improve the accuracy of baseline models for in-the-wild datasets.

A different approach was taken in \cite{okabe_attentive_2018} and \cite{zhu_self-attentive_2018}. The attention models used in these studies replaced the statistical pooling layer of an X-Vector model. The proposed models utilized TDNN to extract features from short frames of the utterances. Attention models were then used to aggregate the features into an utterance-level embedding. The model proposed in \cite{zhu_self-attentive_2018} was evaluated against the NistSRE16 evaluation set \cite{NistSRE_2016} and the proposed model in \cite{okabe_attentive_2018} was evaluated against the VoxCeleb test set \cite{nagrani_VoxCeleb}. Both models showed substantial improvements compared to their baseline models.

In Table \ref{tab:litreview}, we summarize the prominent studies that use attentive models for SR and SER. These studies are grouped based on the task (SR/SER), and then by the type of attention mechanisms used. The table also presents the type of networks used. We also review the evaluation datasets, whether they are in-the-wild, and the type of features used as input to each system, in order to obtain a better picture of the type of data used with existing attention mechanisms for SR and SER. For completeness, we also include the main loss functions used in each study. Lastly we state whether each of the prior works are `early' mechanisms and `fine-grained' or not. We define early attention as those which are applied to the \textit{input} spectrograms prior to processing by the DNN, while fine-grained is defined as the ability of the attention mechanism to consider and focus on each frequency-bin individually.

Based on Table \ref{tab:litreview}, we make the following observations. First, it can be seen that despite attention mechanisms being widely used in SR, their use for SER has been limited. Second, we observe that the majority of attention mechanisms proposed in the literature use the features obtained from DNNs as the information items. Moreover, the queries of the attention models have been originated from a latent layer of the model from which the utterance-level embeddings are retrieved. Generally, DNNs learn to extract a low-dimensional latent representation from the input data without necessarily preserving localization with respect to the input information items. Thus, while the use of the latent layer of a DNN for extracting the query of an attention mechanism can be advantageous due to its reduced number of parameters, high levels of granularity and a localized relationship with respect to the input may not be achieved. As a result, most prior attention-based methods do not possess the ability to attend to highly granular individual input frequency features. Compared to the methods proposed in previous studies, our fine-grained attention model proposed in this paper does \textit{not} rely on embeddings obtained from DNNs, and instead operates on spectrograms extracted from raw audio signals. Hence, the granularity of the attention model can be improved to attend to frequency-level features. 

\section{Method}
\subsection{Overview}
In this section we first describe the individual speech-based information items which our model intends to use for the proposed attention mechanism. Next we present our proposed method, FEFA, which is a special form of general attention specifically designed for being integrated into speech processing pipelines and using spectral information for enhancing the learned audio representations. We then define the kernel used in our model and propose two different kernel types, which we experiment with later in the next Section. Lastly we describe the different approaches for integrating FEFA into existing CNN speech-related pipelines.

\subsection{Proposed Mechanism}

\noindent\textbf{Information Items.} 
The spectrogram representation of speech signals is a frequently used input among deep learning models that exploit CNN architectures. While the number of frequency-bins may vary in different studies, the overall approach in calculating and using spectrogram representations are quite similar. The spectrogram representation of an utterance is obtained using Short-time Fourier Transform (STFT) as follows 
\begin{equation}
  S\{x(t)\}(\tau,\omega) = \int_{-\infty}^{\infty} x(t) M(t-\tau) e^{-i \omega t} \, d t  , 
  \label{eq:stft_1}
\end{equation}
where $x(t)$ serves as the signal amplitude at a given time $t$. $M(t-\tau)$ is a mask function applied over the signal to enforce the time window of the STFT as well as to extract the phase information of the signal. $\omega$ represents the frequency band around which the STFT is performed. The calculated STFT values for different frequency bands $\omega$, are then squared over a time window of \textbf{$\tau$}, where the default value is set to 25 \textit{ms}, as follows:
\begin{equation}
  S'(x(t),\omega_i) =  |S\{x(t)\}(25, \omega_i)|^{2} .
  \label{eq:stft_2}
\end{equation}

The value of 25 \textit{ms} is selected for $\tau$ through a series of experiments. Values lower than 25 \textit{ms} are observed to degrade the performance of DNN models. This may be due to the limited range of frequencies that can be extracted from such small window sizes. For $\tau$ values larger than 25 \textit{ms}, no noticeable improvement is observed in the performance of DNN models, however, the computational load rises exponentially.

Finally the spectrogram representation of the speech signal is obtained by repeating this process for a select number of frequency bands. The frequency bands are generally selected as a hyper-parameter in the form of a set of filters called filter-bank. Each value acquired by function $S'(x(t),\omega_i)$ represents the frequency information of the signal with regards to the frequency band $\omega_i$ at a given time in a 25 \textit{ms} time-window. By having the frequency-bins as the construction blocks of the spectrogram representations, every individual bin can be considered the smallest item carrying information, i.e., \textbf{information item}.

\begin{figure*}[!t]
\centering
\includegraphics[width=0.8\linewidth]{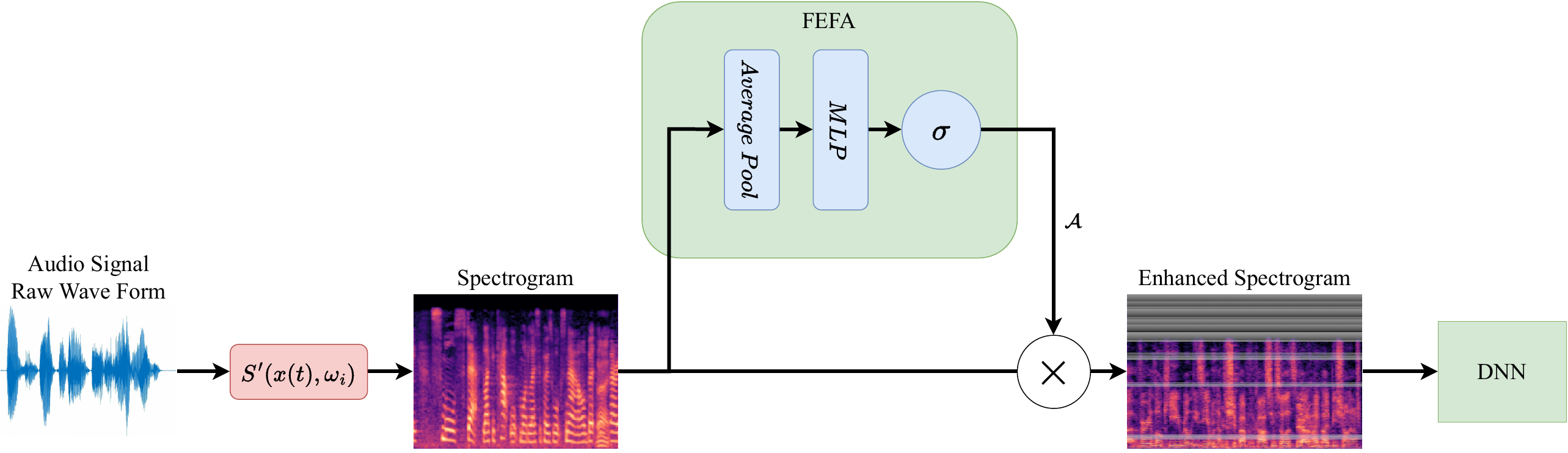} 
\caption{The overview of our proposed model. The model uses the spectrogram representation of the utterance as the information items and the feature set associated with the first layer of the DNN as the source from which to extract the query. The modules inside FEFA consist of an average pooling layer in either Locally Connected (LC) or Fully Connected (FC) configurations.}
\label{fig:ffa_scheme}
\end{figure*}

\noindent\textbf{FEFA.} The fundamental paradigm of a general attention mechanism is the memory-query system. In general attention mechanisms used for audio signals, the memory typically consists of a set of information items, namely DNN embeddings of utterances, while the query is acquired from the layer immediately after the attention module in the DNN. The memory set $M$ is saved in the form of key-value tuples ($k_i$, $x_i$). The first element of the tuple $k_i$ helps with the calculation of the probability factor $p_i$, which indicates the impact of the item in the output of the attention mechanism.
This probability factor is calculated by
\begin{equation}
  p_i(k_i) = \frac{e^{k_i\times W}}{\sum\limits_{j=1}^{|M|} e^{k_j\times W}},
  \label{eq:attention_1}
\end{equation}
where $W$ is a set of trainable weights learned by the attention mechanism. The final output of the attention mechanism, $O$, with respect to the query $q$, is the expected value of items with regards to $x_i$ as follows
\begin{equation}
  O_{q}^{M} = \sum\limits_{i=1}^{|M|} p_i(k_i) \times x_i.
  \label{eq:attention_2}
\end{equation}

With this formulation and by taking into consideration that the source of query is usually the final layer of the DNN, using the frequency-bins as $x_i$ will result in a very large set of weights $W$, which in turn increases the complexity of the attention module. However, by localizing $q$ to individual frequency-bins through changing the source of the query to the first layer of the DNN (illustrated in Figure \ref{fig:ffa_scheme}), $W$ can be divided into multiple smaller groups denoted by $W_i$ each corresponding to a frequency-bin. Thus, the probability-factor $p_i$ can be calculated using the following
\begin{equation}
  p_i = \frac{e^{I(S'(x(t),\omega_i),F))\times W_i}}{\sum\limits_{j=1}^{|M|} e^{I(S'(x(t),\omega_j),F))\times W_j}},
  \label{eq:FFA_1}
\end{equation}
where the weight group $W_i$ can be as small as a single scalar. {Here $I$ is the index of the frequency bin in the input vector which acts as the positional encoding of the frequency bin in the attention mechanism.We use one-hot encoding for positional coding of the frequency-bins.}

Subsequently the output of the attention module, namely the attention map $A$, can be calculated as follows
\begin{equation}
  A = \sum\limits_{i=1}^{|M|} p_i \times S'(x(t),\omega_i),
  \label{eq:FFA_2}
\end{equation}
where $S'$ is the frequency-bin used as information items calculated in Eq. \ref{eq:stft_2} and $|M|$ is number of bins. The attention map acquired from FEFA is then multiplied by the original spectrogram representation of the utterance to enhance its representation for subsequent use by the DNN.

The memory and computational complexities of FEFA are respectively linear ($\Theta(n)$) and quadratic ($\Theta(n^2)$) with regards to the number of frequency-bins $|M|$ in the spectrogram representation. Hence adding multiple layers of FEFA throughout the DNN pipeline does not drastically increase the computational complexity of the model. We perform empirical experiments on the complexity of our proposed method later in Section \ref{result:complexity}. As another advantage of our approach, FEFA can be integrated into various deep learning pipelines without any need to change their architecture as long as they use spectrogram representations of utterances as input. This plug-and-play characteristic of our method makes it desirable, easy to use, and compatible with any DNN pipeline, as demonstrated later in Section \ref{Experiments}.

\begin{figure}[!t]
\centering
\includegraphics[width=0.75\columnwidth]{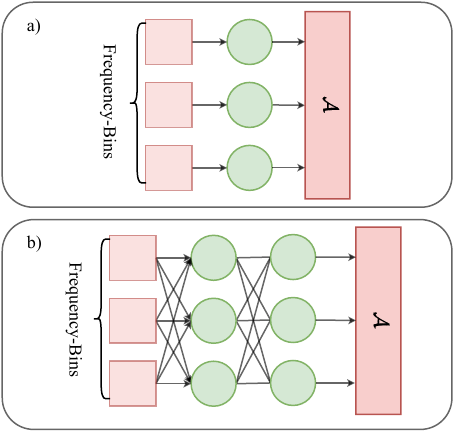} 
\caption{ Tow configurations of the FEFA, (a) Using a single locally connected layer, (b) Using multiple fully connected layers.}
\label{fig:ffa_kernal}
\end{figure}

\noindent \textbf{Kernel Type.}
We use an MLP as the main kernel for FEFA as illustrated in Figure \ref{fig:ffa_scheme}. Accordingly we define two different configurations for this kernel:
\begin{itemize}
    \item Locally Connected (LC);
    \item Fully Connected (FC). 
\end{itemize}
Using a locally connected layer (see Figure \ref{fig:ffa_kernal} (a)) simply assigns a specific weight to each individual frequency-bin. This implies that the weight of the corresponding frequency-bin in the attention map is only determined based on the value and the index of the frequency-bin itself and the query, while any information from other frequency-bins is ignored. In our experiments, we denote this mechanism by FEFA(LC). On the other hand, by using a fully connected layer (see Figure \ref{fig:ffa_kernal} (b)), the model considers all the frequency-bins to determine the weight assigned to each frequency-bin in the final attention map. This is denoted in our experiments by FEFA(FC). In Section \ref{Experiments} we perform extensive experiments to study both kernel types in different settings.

\noindent \textbf{Kernel Integration.}
\label{sec:integration}
As discussed earlier, FEFA (with either kernel type) can be easily integrated into different CNN architectures that learn and process speech representations. It is noteworthy to add that while we believe FEFA would benefit from integration into the first layer of CNNs so that individual salient frequency-bins can be attended to, it is highly feasible to integrate the module prior to every convolution layer of the pipeline. Accordingly, we define two integration methods for FEFA:
\begin{itemize}
\item Single integration in the first layer;
\item Multiple integrations, one prior to every layer.
\end{itemize}
The definition of the second category is inspired by \cite{yadav2020frequency} where an attention mechanism was integrated prior to every layer of CNN pipelines, and it was discussed that employing this strategy benefits the learned representations. We skip the integration of FEFA at the last layer of the pipeline as it is similar to the conventional attention models which have been studied in prior work \cite{okabe_attentive_2018, bian2019self, cai2018analysis}. In Section \ref{restuls:performance} we perform extensive experiments on both strategies (Single vs. Multi) and demonstrate that FEFA is best suited for a single-integration into different CNNs. This further adds to the ease-of-use of our model and the need for very little added computations for learning enhanced audio representations.

\section{Experiments}
\label{Experiments}

In order to extensively evaluate FEFA along with its different kernel types and integration strategies, we select two tasks on speaker representation learning, namely SR and SER. Additionally, we include spoken digit recognition as the third and a more general, yet simple task, on which various aspects of our method can be easily studied. As mentioned earlier, FEFA can be integrated into different CNN backbones that take spectrogram representations of utterances as input. Hence we use a select number of prominent CNN architectures that are commonly used for these tasks as our benchmarks. In the following subsections, we introduce the datasets used in our experiments, implementation details regarding FEFA, as well as the details of the backbone networks used in which we integrate FEFA.

\subsection{Datasets}
Here we describe the three datasets used for our experiments. For the task of speaker recognition, we experiment on the widely used VoxCeleb dataset \cite{chung_voxceleb}, while we use the IEMOCAP dataset \cite{busso2008iemocap} for the task of speaker emotion recognition. Lastly we use the Free Spoken Digit Dataset (FSDD) \cite{amnist} for the spoken digit recognition task. Following we provide further details on each of these datasets.

\begin{table}[!t]
    \caption{Number of instances for training, validation, and test splits for VoxCeleb, IEMOCAP (for each fold), and FSDD datasets.}
    \label{tab:splits}
    \centering
    \begin{tabular}{l|c|c|c}
    \hline
        Dataset & Training & Validation & Test\\ \hline \hline
        VoxCeleb & 1,092,009 & -- & 4,874 \\ \hline
        \multirow{2}{*}{IEMOCAP} & 4800  & -- & 1200  \\
                & (1200 per class) & & (300 per class) \\ \hline
        \multirow{2}{*}{FSDD} & 1680 & 240  & 480  \\ 
             &  (168 per class) & (24 per class) & (48 per class) \\
        \hline
    \end{tabular}
    
\end{table}

\noindent \textbf{VoxCeleb.} For the SR task we perform our evaluations using the large and widely used \textit{in-the-wild} VoxCeleb dataset \cite{chung_voxceleb}. This dataset includes voices from more than 6,000 individuals. The utterances are captured in uncontrolled conditions such as interviews published in open-source media. The VoxCeleb dataset is available in two versions, VoxCeleb1 which is used more commonly for evaluation, and VexCeleb2 which is used solely for training purposes. In this experiment we follow the common practice and use the VoxCeleb2 dataset with nearly 1.2 million utterances for training our model and VoxCeleb1 test set for evaluation. Similar to prior work such as \cite{bian2019self, okabe_attentive_2018, kye2021supervised, india2021attention_speaker, jung2021attention_speaker, safari2019self, wang2020attention_speaker, zhou2019attention_speaker}, and \cite{yadav2020frequency}, we use the entirety of the VoxCeleb2 dataset (not VoxCeleb1) for training. We then test the solution on the 38,000 trial pairs (4,874 unique utterances selected by the authors of VoxCeleb, from VoxCeleb1 dataset) spoken by 40 individuals. The trial pairs are labeled by ``1'' and ``0'', indicating whether both utterances in the pair are spoken by the same person or not. The number of pairs labeled as 0 or 1 is completely balanced. The details of the distributions of the instances between different splits are mentioned in Table \ref{tab:splits}.

\noindent \textbf{IEMOCAP.} For the task of SER, we use the IEMOCAP dataset \cite{busso2008iemocap}, which is widely used in the field. This dataset is a multi-modal emotion recognition dataset including speech recordings, videos, and motion capture. The dataset contains 12 hours of prompted and improvised dialogue performed by 10 actors. The audio recordings of the dataset are divided into short utterances each containing one sentence. Each utterance is then scored by several people to determine the category of emotion conveyed by the utterance. Four emotion categories, namely \textit{Sadness}, \textit{Happiness}, \textit{Angry}, and \textit{Neutral}, for a total of 6 thousand utterances, are used. We use k-fold cross validation with $k=5$ for training and evaluating the backbone networks with and without FEFA. In order to maintain the balance between the classes of dataset we select equal number of utterances for each class both in the training split and test split during each run of cross validation process. This approach has been used in literature such as \cite{tarantino2019self, ramet2018context, mirsamadi2017automatic}, and \cite{chernykh2017emotion}. Table \ref{tab:splits} shows the number of instances used for training and testing in each fold as well as the number of instances for each class in the splits.

\noindent \textbf{FSDD.} The free spoken digit dataset \cite{amnist} is an open source dataset, consisting of recordings from 6 individuals, each uttering a single digit 40 times resulting in a total number of utterances equal to 2400 utterances. Most of the utterances have a duration of approximately 1 second with the longest utterance being 1.5 seconds. We have zero-padded the utterances to a fixed length of 1.5 seconds to maintain consistency throughout the dataset. These utterances are labeled from zero to nine corresponding to the digit spoken in the utterance. Unlike the other datasets used in this paper, there are no standard procedures for splitting FSDD into training and test sets. Therefore we select the standard method of splitting the dataset into 3 sets of training, validation, and test sets with the respective ratios of 70\%, 10\%, and 20\% while maintaining a complete balance between the classes of utterances. The number of instances used for training,validation, and testing splits of the FSDD dataset are also presented in Table \ref{tab:splits}.

\subsection{Data Preparation}
For data preparation, we extract spectrogram representations of the utterances {using 512 filter bands which are generated using ``hann'' function. The spectrograms are extracted using STFT on overlapping time windows with size of 25 \textit{ms} with a stride of 10 \textit{ms}}. This results in spectral images of size $257\times T$ {where T is equal to the number of strides required for the time window $\tau$ to cover the entire utterance and in turn is equal to the dimensionality of the spectrograms in the time axis.} We use 257 frequency-bins as it is standard practice in \cite{xie2019utterance, chung_voxceleb, nagrani_VoxCeleb}. 

\subsection{FEFA Modules}

For our proposed module, we use the average pooling layer with the input size of $257\times T$, and output size of $257$. In the LC configuration of FEFA the output of the average pooling layer is given to an LC layer with the output shape of $257\times 1$. In the FC configuration of FEFA the output of the average pooling layer is given to two FC layers each with $257$ hidden units which results in an output with a shape of $257 \times 1$. The output of either LC or FC layers is then passed through a sigmoid activation function that generates a $257\times 1$ attention map.

\subsection{Backbone Models}
{ FEFA can be attached to any backbone network that uses spectrograms as input. This makes FEFA a simple standalone module that is independent of the DNN architectures used for performing speaker representation learning.} We select three popular CNN architectures as the backbones in which to integrate FEFA and its different configurations. The choice of these backbones is based on a survey of the literature to identify the common architectures used for audio-representation learning \cite{hajavi2019deep, xie2019utterance, chung_voxceleb, nagrani_VoxCeleb,  gulati2020conformer, bian2019self, kye2021supervised, india2021attention_speaker, xie2019speech, cai2018speaker,  yadav2020frequency, vydana2017residual, celano2021resnet, mamyrbayev2020end,  hajibabaei2018unified, hajavi2021siamese, tedd2021fluent}. The models proposed in these studies use architectures based on the selected backbone models and they are often compared to these backbone networks in their experiments.

The first backbone CNN used in our experiments is the VGG-based model proposed in \cite{nagrani_VoxCeleb}, which consists of 5 convolution layers accompanied by 3 maxpooling layers. The utterance-level aggregation is done using a global-average-pooling and the final embedding is acquired using an FC layer with ReLU activation.

The second network that we use in this experiment is the ResNet-based model proposed in \cite{xie2019utterance}. This model consists of 35 convolution layers used in the form of residual blocks. The shortcuts integrated in residual blocks help the model convey the learning gradients throughout the pipeline of the model more easily, which in turn aids the model to learn faster and more effectively. This also enables the model to provide better queries for FEFA. The complete details about the hyper-parameters of this backbone can be found in \cite{xie2019utterance}.

The final network used in our experiments is SEResNet. Similar to the ResNet model, the SEResNet consists of residual blocks. The formation of blocks and number of parameters used in the SEResNet is similar to a ResNet with an addition of a Squeeze-and-Excitation (SE) module \cite{hu2018squeeze}. The SE module uses a global pooling layer to extract channel information inside the residual blocks. The channel information is then projected onto a latent space using 2 FC layers, a ReLU activation function, and a Sigmoid activation function. The resulting representation is then multiplied across channels of the ID block. 

\subsection{Training}
We use the cyclical learning rate proposed in \cite{smith2017cyclical} for training the backbone networks with the added FEFA module. This technique helps the model to achieve a better convergence by changing the learning rate periodically and preventing the model from getting trapped in local minima. An initial learning rate of $10^{-4}$ is used. Throughout the entire training process, the models are trained using a single Nvidia Titan RTX (24 GB vRAM) GPU. We use a batch-size of 64 and Adam optimizer \cite{kingma2015adam} for training. The values selected for the hyperparameters are obtained empirically. The training time of the backbone models do not experience any noticeable change after attaching the FEFA modules. This may be due to the fact that the number of parameters of the modules are considerably low, especially in comparison to the total number parameters in the backbone models. This shows that our proposed solution can be added to backbone networks without any considerable additional costs in terms of training time or energy consumption.

\begin{figure}[!t]
\centering
\includegraphics[width=.7\columnwidth]{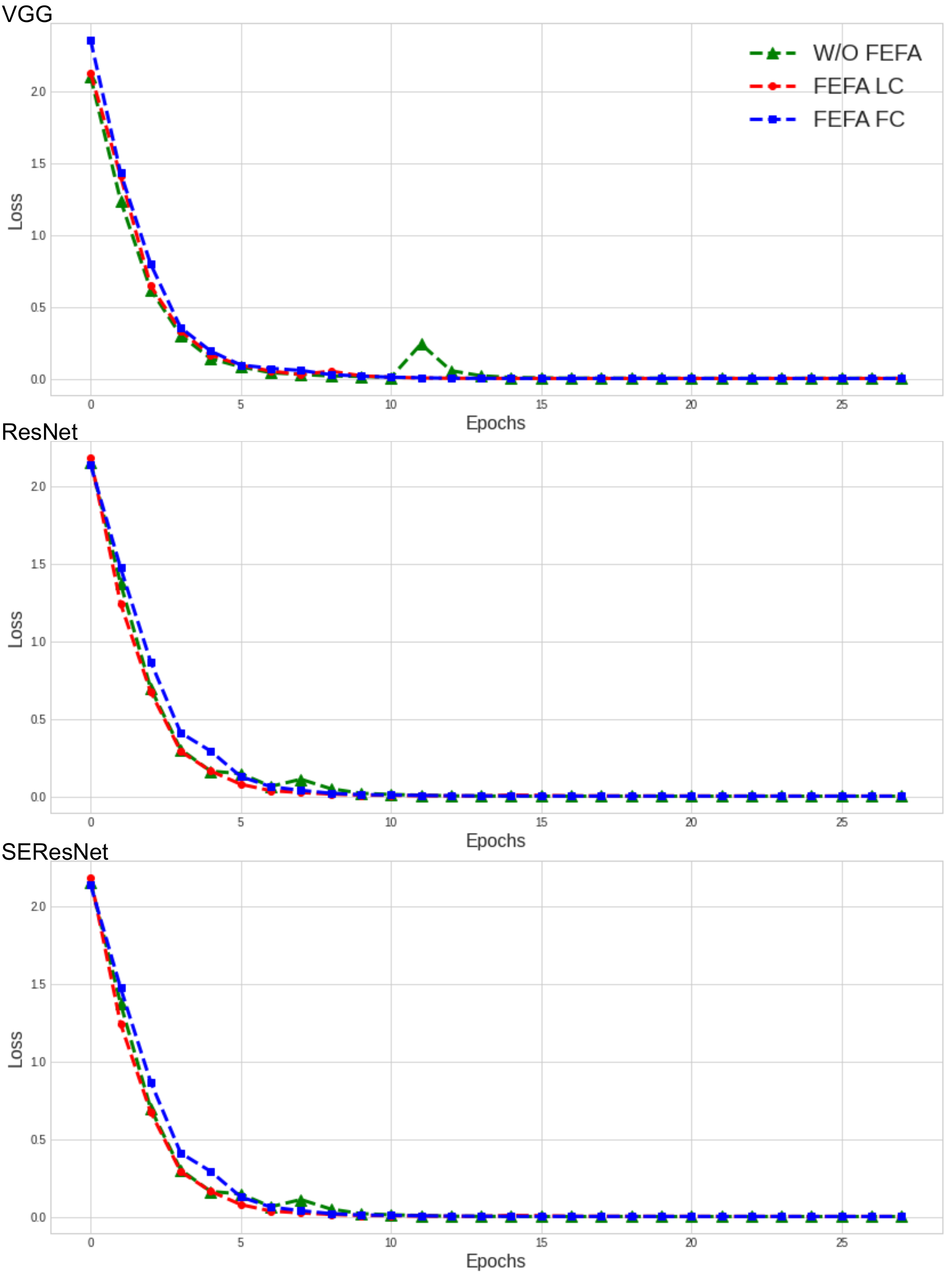} 
\caption{The training curves of the backbones with and without integration of different configurations of FEFA.}
\label{fig:training_curves}
\end{figure}

\begin{table}[!t]
\small
\caption{Results for the spoken digit recognition task using different backbones equipped with different integration and configurations of FEFA.}
\resizebox{1\columnwidth}{!}{%
\begin{tabular}{l|l|c|c|c}
\hline
Model    & Backbone   & Integration           & $\uparrow$ACC (\%)             & $\uparrow$$\Delta$ACC (\%)     \\ \hline \hline
 
          & VGG \cite{lee2021acav100m}       & --    & 65.4     & -- \\ 
 w/o FEFA & ResNet                           & --    & 74.5     & --     \\ 
          & SEResNet                         & --    & 75.8     & --     \\ \hline
          & VGG                              & Single     & 71.2     & +8.87     \\
          & VGG                              & Multi      & 69.1     & +5.66     \\
 FEFA(LC)  & ResNet                           & Single     & 76.7     & +2.95     \\
          & ResNet                           & Multi      & 75.3     & +1.07     \\
          & SEResNet                         & Single     & 76.7     & +1.18     \\
          & SEResNet                         & Multi      & 76.4     & +0.79     \\ \hline
          & VGG                              & Single     & 75.0     & +14.67     \\
          & VGG                              & Multi      & 73.4     & +12.32     \\
 FEFA(FC)  & ResNet                           & Single     & 77.9     & +4.56     \\
          & ResNet                           & Multi      & 76.8     & +3.08     \\
          & SEResNet                         & Single     & 79.9     & +5.40     \\
          & SEResNet                         & Multi      & 78.5     & +3.56     \\ \hline
\end{tabular}
}
\label{tab:fsdd_results}

\end{table}

\begin{table*}[!th]
\footnotesize
\centering
\caption{Results for speaker verification using the standard VoxCeleb1 test set. TAP: Temporal Average Pooling; AP: Attentive Pooling. *The original work in \cite{yadav2020frequency} reports the EER values with ArcSoftmax loss. However, in order to maintain consistency and provide a fair comparison, we retrained the model with Softmax.}
\label{tab:resultsSR}
\begin{tabular}{l|c|c|c|c|c|c|c}
\hline
study                                             & Backbone      &  Loss      & Attention              & Dims      & Aggregation      & Training Set      & EER (\%)\\ \hline \hline
Narani et al. \cite{nagrani_VoxCeleb}             & i-Vector+PLDA & Softmax & --                     & --        & --               & VoxCeleb1         & 8.8     \\
Narani et al. \cite{nagrani_VoxCeleb}             & VGG           & Softmax  & --                     & 1024      & TAP              & VoxCeleb1         & 7.8     \\
Hajibabai et. \cite{hajibabaei2018unified}        & ResNet20      & Softmax  & --                     & 128       & TAP              & VoxCeleb1         & 4.30    \\
Chung et al.  \cite{chung_voxceleb}               & ResNet50      & Softmax  & --                     & 512       & TAP              & VoxCeleb2         & 3.95    \\
Xie et al. \cite{xie2019utterance}                & Thin-ResNet   & Softmax  & --                     & 512       & TAP              & VoxCeleb2         & 10.48   \\
Xie et al. \cite{xie2019utterance}                & Thin-ResNet   & Softmax  & --                     & 512       & GhostVLAD        & VoxCeleb2         & 3.22    \\
Okabe et al. \cite{okabe_attentive_2018}          & x-Vector      & Softmax  & Soft Attention         & 1500      & AP              & VoxCeleb1         & 3.85    \\
Bian et al. \cite{bian2019self}                   & ResNet50      & Softmax  & Self Attention         & 512       & AP                          & VoxCeleb2          & 5.4     \\
India et al. \cite{india2021attention_speaker}    & VGG           & Softmax  & Multi-head      Attention        & 1024      & AP        & VoxCeleb2          & 3.19    \\
Kye et al. \cite{kye2021supervised}               & Resnet34      & Softmax  & Supervised Attention   & 256       & AP        & VoxCeleb2          & 4.75    \\
Cai et al. \cite{cai2018speaker}                  & ResNet34      & Softmax  & Self Attention                     & 128       & AP              & VoxCeleb1         & 4.40    \\
Yadav et al.* \cite{yadav2020frequency}      & Thin-ResNet         & Softmax*  & CBAM                   & 512       & GhostVLAD        & VoxCeleb2          & 3.10    \\ 
\hline
Proposed                                          & VGG       & Softmax       & FEFA(LC)               & 1024      & TAP              & VoxCeleb2           & 7.43   \\
Proposed                                          & VGG        & Softmax      & FEFA(FC)               & 1024      & TAP              & VoxCeleb2           & 7.10   \\
Proposed                                          & SEResNet   & Softmax      & FEFA(LC)               & 512       & TAP              & VoxCeleb2           & 3.48   \\
Proposed                                          & SEResNet   & Softmax      & FEFA(FC)               & 512       & TAP              & VoxCeleb2           & 3.43   \\ 
Proposed                                          & Thin-ResNet   & Softmax   & FEFA(LC)               & 512       & TAP              & VoxCeleb2           & 5.40   \\
Proposed                                          & Thin-ResNet   & Softmax   & FEFA(FC)               & 512       & TAP              & VoxCeleb2           & 5.32   \\
Proposed                                          & Thin-ResNet  & Softmax    & FEFA(LC)               & 512       & GhostVLAD        & VoxCeleb2           & 2.78   \\
Proposed                                          & Thin-ResNet    & Softmax  & FEFA(FC)               & 512       & GhostVLAD        & VoxCeleb2           & \textbf{2.61}   \\ \bottomrule
\end{tabular}

\end{table*}

\begin{table}[!t]
\label{tab:benchmarkSR}
\caption{Results for different integrations of FEFA on different backbone networks for SR.}
\centering
\resizebox{1\columnwidth}{!}{%
\begin{tabular}{l|l|c|c|c}
\hline
 Model    & Backbone                         & Integration      &  EER (\%)  & $\Delta$EER (\%)     \\ \hline \hline
 
           & VGG \cite{lee2021acav100m}       & --    & \underline{7.8}  & --      \\ 
 Benchmark & ResNet \cite{xie2019utterance}   & --    & \underline{3.22} & --      \\ 
           & SEResNet                         & --    & \underline{4.81} & --      \\ \hline
           & VGG                              & Single     & 7.43             & -4.7    \\
           & VGG                              & Multi     & 7.62             & -2.3    \\
 FEFA(LC)  & ResNet                           & Single     & 2.78             & -13.6   \\
           & ResNet                           & Multi     & 3.08             & -4.3    \\
           & SEResNet                         & Single     & 3.48             & -25.6   \\
           & SEResNet                         & Multi     & 3.58             & -25.5   \\ \hline
           & VGG                              & Single     & 7.10             & -8.9    \\
           & VGG                              & Multi     & 7.26             & -6.9    \\
 FEFA(FC)  & ResNet                           & Single     & 2.61             & -18.9   \\
           & ResNet                           & Multi     & 2.91             & -9.6    \\
           & SEResNet                         & Single     & 3.43             & -28.6   \\
           & SEResNet                         & Multi     & 3.47             & -27.8   \\ \hline
\end{tabular}
}
\end{table}

\section{Results}
In this section we provide the results of our experiments on the aforementioned datasets. First we illustrate the impact of integrating FEFA into different backbones on the training process. We then present the results of different backbones with different FEFA architectures in comparison to a large number of benchmarks. Through the following section, we experiment with added synthetic noise in order to test the robustness of FEFA in comparison to other models. Lastly, drawing on the results obtained in this section, we provide a discussion on the differences between FEFA and other forms of attention.

\subsection{Impact on Training}
In order to fully understand the effect of FEFA on training the backbone networks, we compare the training curves of the backbone networks with and without integration of FEFA with different configurations. This comparison is presented in Figure \ref{fig:training_curves} for FSDD. Here, we observe that the backbone networks equipped with FEFA not only don't experience any negative impact from the added modules, but in some cases even have smoother training curves and converge faster compared to the backbone networks without the added FEFA.

\subsection{Performance}
\label{restuls:performance}
We use classification accuracy to measure the performance of backbone models equipped with different integrations and configurations of FEFA on the task of spoken digit recognition. The results, as presented in Table \ref{tab:fsdd_results}, show that when the FEFA modules are integrated into the DNN pipeline, the performance of the model is improved. The highest performance gain is achieved by adding the fully connected configuration of FEFA into the VGG pipeline, where a relative $14\%$ boost is obtained compared to the baseline. This is while improvements of $4\%$ and $5\%$ are achieved by adding the same configurations to the ResNet and SEResNet backbones.

\begin{table*}[!t]
\caption{Results of SER on IEMOCAP Dataset. UA: Unweighed Accuracy.}
\small
\centering
\begin{tabular}{l|l|c|c|c|c|c}
 \hline

Study     & Model      & Attention   &  Feature Type & Use Other Modalities &  UA & F1 \\ \hline \hline

Chernykh et al. \cite{chernykh2017emotion}     & RNN             & --              & MFCC       & Yes & 54      &  --  \\
Ramet et al. \cite{ramet2018context}           & RNN             & Self Attention  & MFCC+LLD   & No  & 59.6    &  --  \\
Mirsamadi et al. \cite{mirsamadi2017automatic} & RNN             & Local Attention & MFCC+LLD   & No  & 58.8    &  --  \\
Tarantino et al. \cite{tarantino2019self}      & Transformer     & Self Attention  & MFCC+LLD   & No  & 63.8    &  --  \\ \hline
Proposed                                       & VGG             & FEFA(LC)        & Spect 256  & No  & 56.70   &  56.83  \\
Proposed                                       & VGG             & FEFA(FC)        & Spect 256  & No  & 58.23   &  58.41  \\
Proposed                                       & SEResNet        & FEFA(LC)        & Spect 256  & No  & 62.28   &  62.43  \\
Proposed                                       & SEResNet        & FEFA(FC)        & Spect 256  & No  & 63.92   &  64.02  \\ 
Proposed                                       & ResNet     & FEFA(LC)        & Spect 256  & No  & 62.32        &  62.46  \\
Proposed                                       & ResNet     & FEFA(FC)        & Spect 256  & No  & \textbf{63.97}& 64.09  \\ \bottomrule
\end{tabular}
\label{tab:ser_results}
\end{table*}

\begin{table}[!t]
\caption{Results for different integrations of FEFA on different backbone networks for SER.}
\small
\centering
\begin{tabular}{l|l|c|c|c}
 \hline
 Model    & Backbone   & Integration         & UA (\%)             & $\Delta$UA (\%)     \\ \hline \hline
          & VGG                              & --    & 52.48     & --     \\ 
 Benchmark& ResNet                           & --    & 59.72     & --     \\ 
          & SEResNet                         & --    & 59.82     & --     \\ \hline
          & VGG                              & Single     & 56.70     & +8.04     \\
          & VGG                              & Multi      & 55.36     & +5.48     \\
 FEFA(LC) & ResNet                          & Single     & 62.32     & +4.36     \\
          & ResNet                           & Multi      & 61.57     & +3.09     \\
          & SEResNet                         & Single     & 62.28     & +4.11     \\
          & SEResNet                         & Multi      & 61.63     & +3.25     \\ \hline
          & VGG                              & Single     & 58.23     & +10.95     \\
          & VGG                              & Multi      & 57.86     & +10.25     \\
 FEFA(FC) & ResNet                           & Single     & 63.97     & +7.11     \\
          & ResNet                           & Multi      & 63.63     & +6.54     \\
          & SEResNet                         & Single     & 63.92     & +6.85     \\
          & SEResNet                         & Multi      & 63.85     & +6.73     \\ \hline
\end{tabular}
\label{tab:ser_results_benchmark}
\end{table}

For evaluating the networks in the SR domain, we use the commonly used metric of Equal Error Rate (EER). This metric is the error threshold of the model at which the number of false positive errors is equal to the number of false negative errors. Table \ref{tab:resultsSR} presents the results as well as the performance gain achieved by adding different configurations of FEFA to different backbones. The results show that when FEFA is integrated into simple CNN backbones, the model's performance is boosted, outperforming the respective baseline. In particular, we observe that by integrating either configurations of FEFA (LC and FC) into a Thin-ResNet followed by GhostVLAD aggregation, we outperform other works in the area. We should note that the original work in \cite{yadav2020frequency} reports the EER values with ArcSoftmax loss. However, in order to maintain consistency and provide a fair comparison, we retrained their model with Softmax loss. Table \ref{tab:benchmarkSR} presents further experiments on FEFA for SR where we evaluate the performance of different both integration strategies discussed in Section \ref{sec:integration}. It is observed that while both approaches improve the performance by considerable margins, the single integration strategy provides better results. We hypothesize that this could be due to the mixing effect experienced by the individual frequency-bins as they are processed through the network. As a result of this mixing of bins, the individual values attended by FEFA are in fact a mixture of several frequency-bins, and thus the initial goal of providing fine-grained attention on individual frequency-bins is less achieved.

As discussed, we also perform experiments on SER. In these experiments, we employ the commonly used unweighted accuracy (UA) {and F1-Macro score} as evaluation metrics. The F1-Macro score is used for this dataset for future comparisons should the unbalanced version of the dataset be used. On the other hand, for SR and spoken digit recognition, the datasets only have balanced versions and their evaluations are not affected by unbalanced data. To comply with the common practice \cite{chernykh2017emotion, ramet2018context,mirsamadi2017automatic, tarantino2019self} in using the IEMOCAP speech emotion dataset, we perform a k-fold cross validation for evaluating our solution. Given that in many recent works for emotion recognition from speech, VGG, ResNet, and SEResNet architectures have frequently been used for speech representation learning \cite{yenigalla2018speech,kim2017deep}, we utilize the same backbone networks for evaluating the impact of FEFA. This also enables us to compare our results to SR task performed earlier and provide a more consistent analysis of the results. Table \ref{tab:ser_results} shows the results of evaluating the FEFA model for the task of SER. It is evident by the results that the performance of different backbones is boosted by adding FEFA, with ResNet + FEFA with the FC integration outperforming all other works in the area. Moreover, in Table \ref{tab:ser_results_benchmark} we experiment with different integration strategies for FEFA. We observe that much like the case for SR, while both approaches result in a performance boost, better results are achieved when a single integration of FEFA is performed.

\begin{figure*}[!t]
\centering
\includegraphics[width=.7\linewidth]{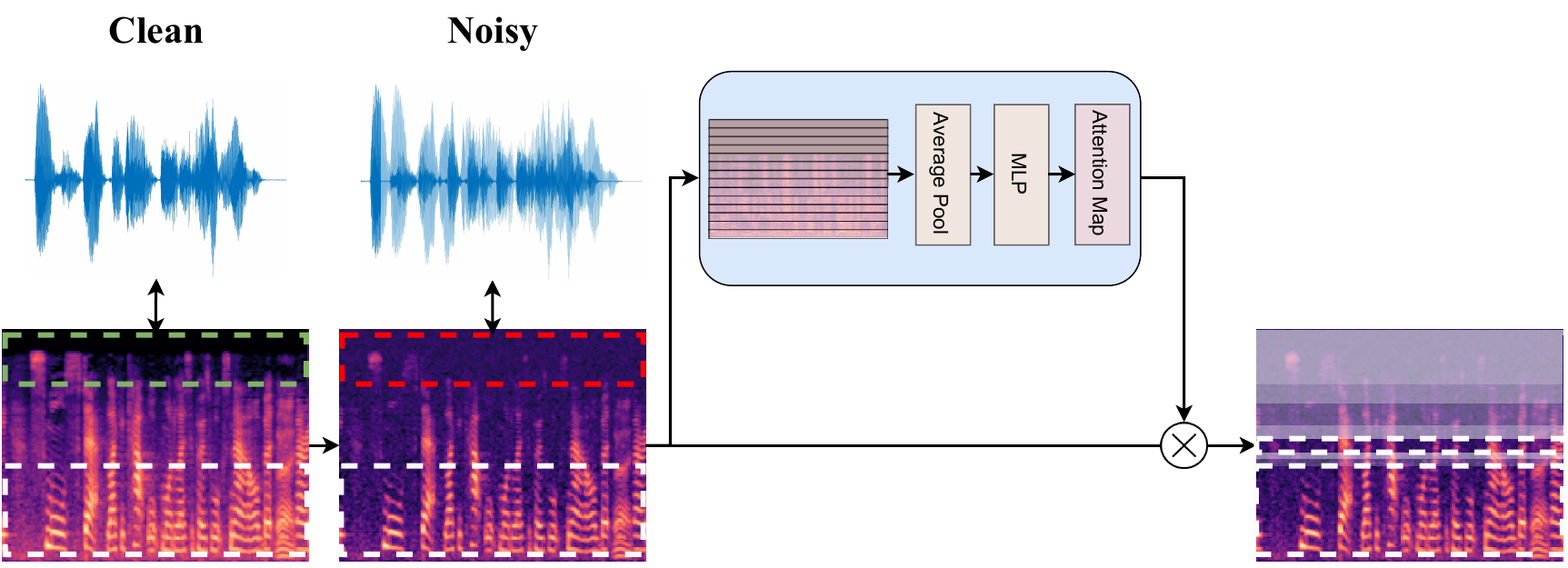}
\caption{Overview of the robustness test against adding synthetic noise to input utterances.}
\label{fig:spectrogram_samples}
\end{figure*}

\begin{figure}[!t]
\centering
\includegraphics[width=1\columnwidth]{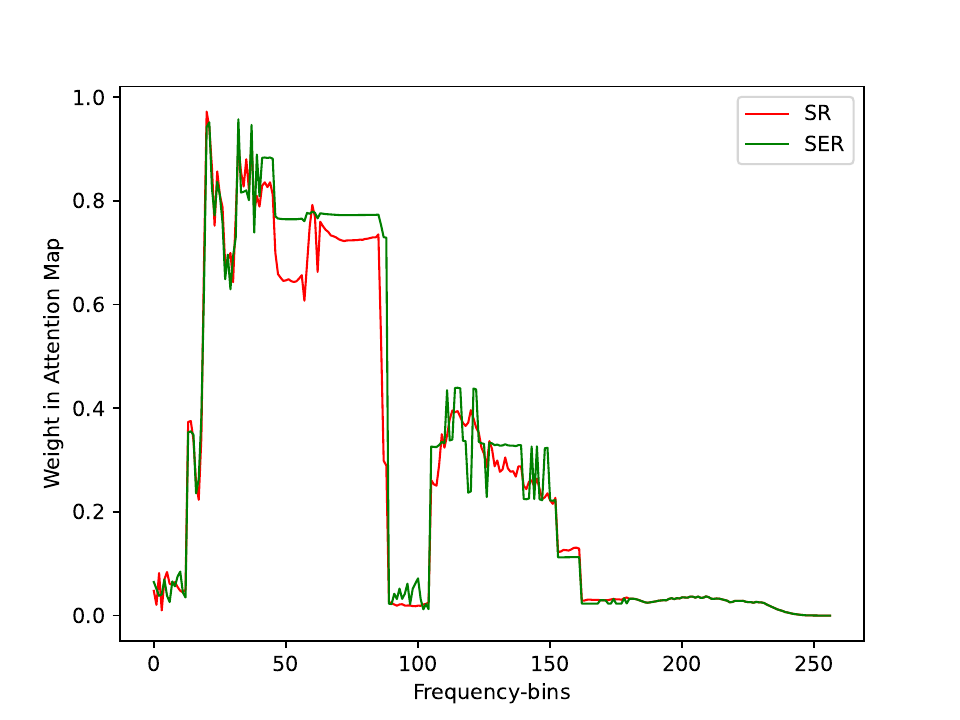} 
\caption{An illustration of generated attention maps for SR and SER.}
\label{fig:attention_map}
\end{figure}

\subsection{Robustness to Noise}
Given that FEFA has been designed to focus on the most salient frequency-bins at the input layer of CNN backbones, we anticipate that by integration of FEFA, robustness to different intensities and types of noise would be enhanced. In order to test this hypothesis, we evaluate the performance of FEFA against different types and levels of noise. To do so, a controlled level of synthetic noise is added to the test utterances used for SR. The model with the best performance from Table \ref{tab:resultsSR} Thin-ResNet + GhostVlad, is selected and tested with the noisy utterances with and without FEFA. It is important to note that the model has not been retrained, and noise is only added to the test samples. The added noise is selected from both Gaussian and uniform distributions. 

\begin{table}[!t]
\small
\caption{Robustness experiment results for SR. We compare FEFA to the state-of-the-art model GhostVlad \cite{xie2019utterance} with and without FEFA.}
\centering
\begin{tabular}{l|l c c c}

\hline
Noise & Model  &  SNR  &  EER (\%) &  $\Delta$EER (\%)\\ 
\hline\hline
       & w/o FEFA & 100db  & 3.40 & +5.5 \\ 
       & w/o FEFA & 50db  & 3.85  & +19.5 \\ 
Normal & w/o FEFA & 20db  & 4.82 & +49.6  \\ \cline{2-5}
       & + FEFA   & 100db  & 2.61 & 0 \\ 
       & + FEFA   & 50db  & 2.78  & +6.5 \\ 
       & + FEFA   & 20db  & 3.10 & +18.7  \\ \hline
        & w/o FEFA & 100db  & 3.32  & +3.1\\ 
        & w/o FEFA & 50db  & 3.48 & +8.0  \\ 
Uniform & w/o FEFA & 20db  & 3.96 & +22.9 \\ \cline{2-5}
        & + FEFA   & 100db  & 2.61 & 0 \\ 
        & + FEFA   & 50db  & 2.80 & +7.2 \\ 
        & + FEFA   & 20db  & 3.10 & +18.7 \\ 

\hline
\end{tabular}
\label{ROBUSTRESULTS}
\end{table} 

Figure \ref{fig:spectrogram_samples} presents the effect of FEFA on a sample spectral representation of a noisy utterance before being fed to a backbone network. The noisy utterance is depicted in the figure with areas most affected by noise highlighted by a red dotted rectangle. This area compared to the clean part of the utterance (highlighted by the green dotted rectangle) contains artifact noise which will negatively impact the performance. The rightmost spectrogram represents the final spectrogram enhanced by FEFA with the attended areas by FEFA shown with the grey mask. This spectrogram has been generated after applying the attention map obtained from FEFA onto the noisy input spectrogram during the experiments. The areas greyed out in the spectral representation show the frequency-bins with the least contribution according to FEFA. Hence by focusing on the other frequency-bins, FEFA decreases the effect of artifact noise in the final learned representation. The results of this test are reported in Table \ref{ROBUSTRESULTS}. The study is performed using 3 levels of synthetic noise with signal-to-noise ratios (SNR) of 20db, 50db, and 100db. As shown by the results, while the performance of the backbone network without FEFA is considerably affected by the added noise, the model with the FEFA mechanism stays considerably more stable.

\subsection{Generated Attention Maps}
Figure \ref{fig:attention_map} shows the final attention weights learned for the tasks of SR and SER. Here we observe that as initially hypothesized, different frequency-bins carry different amounts of importance toward the final task. Specifically, it is demonstrated that the majority of the information for these two tasks lies in the range of frequency-bins 10 to 90, which correspond to 288 Hz to 2.8 KHz. The second important range can be seen as bins 106 to 162, corresponding to 3.3 KHz to 5.2 KHz. Interestingly, it is shown that the other bins have very little to no importance for SR and SER, and thus can be discarded to obtain smaller models and faster inference times.

\begin{figure}[!t]
\centering
\includegraphics[width=1\columnwidth]{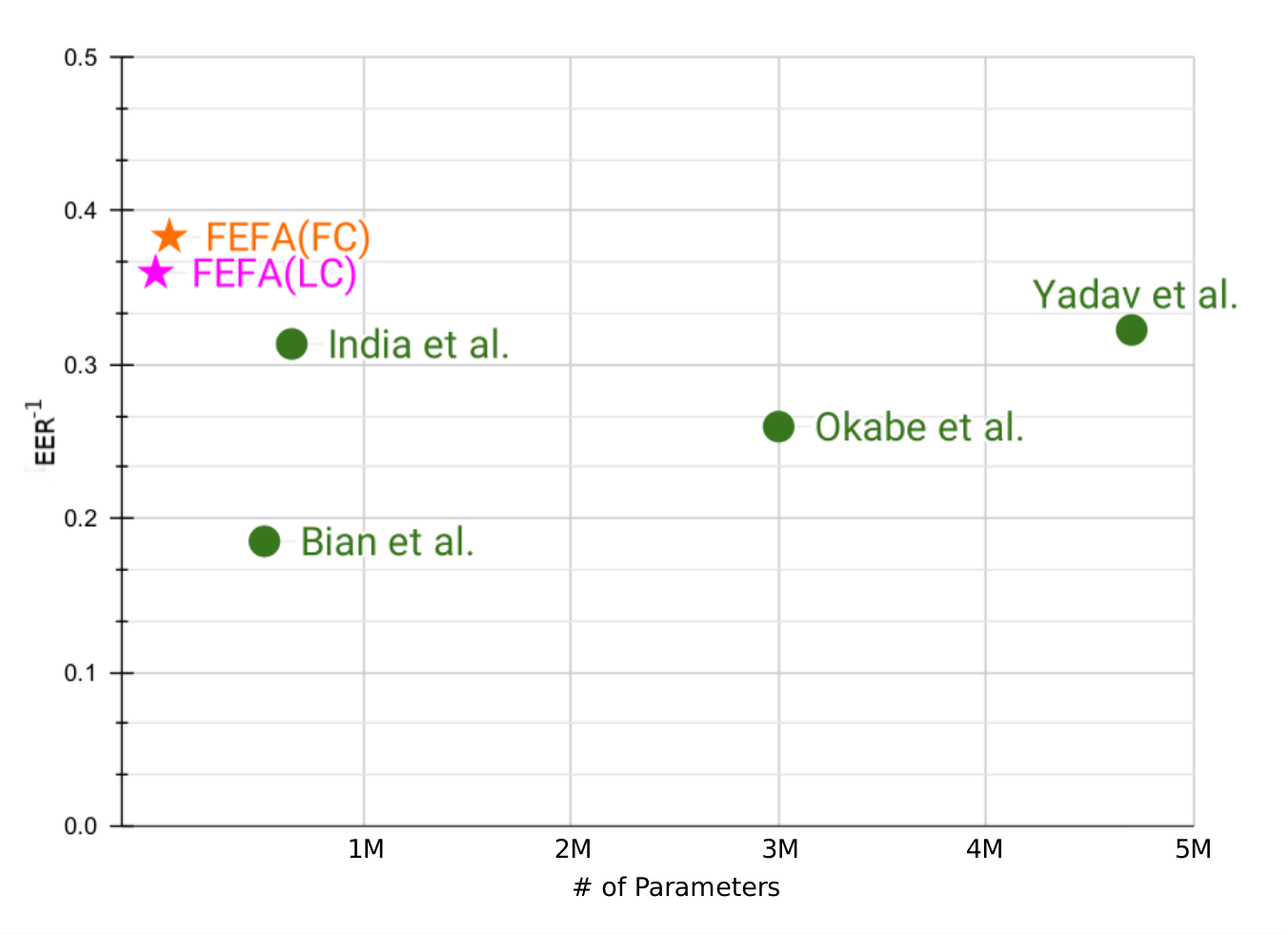} 
\caption{Performance vs. number of parameters for different attention-based models in SR.}
\label{fig:parameters_compare}
\end{figure}

\subsection{Number of Parameters}
\label{result:complexity}
Figure \ref{fig:parameters_compare} portraits a comparison between performance of the two configurations of FEFA and that of the other forms of attention used in SR with respect to the \textit{number of parameters} used. In the figure, the two configurations of FEFA are depicted by stars and the other attention mechanisms are shown by circles. We modify the metric for measuring the performance of the models and use the new metric $EER^{-1}$ so that we can show the better performing models higher in the vertical axis. As shown in the figure, both FEFA models, although using fewer number of parameters, outperform other forms of attention. Naturally, the number of parameters in the FC configuration of FEFA is slightly higher than the LC configuration. However, even in the case of the FC configuration, our module has considerably fewer parameters compared to other forms of attention.

\subsection{Discussion and Comparison to Other Forms of Attention}
As shown in Table \ref{tab:resultsSR}, CNN models plus general forms of attention such as self-attention and soft-attention are outperformed by FEFA integrated into similar backbone networks. Our approach shows a clear performance enhancement over the classical attention mechanisms as such attention models only attend to parts of the \textit{latent representation} that correspond to large areas in the input utterance spectrogram. Hence they fail to focus on very small yet salient frequency-level features that are often crucial in speech-related tasks.

While a number of attempts have been made to achieve different levels of granularity with attention mechanisms, existing attention models do not achieve a fine-grained solution. The model proposed in \cite{li2019area} achieves varying degrees of granularity by creating different combinations of neighboring information items. However, the information items used in the combinations are embeddings already extracted by the DNN, limiting the level of granularity based on the resolution of the latent representation achieved by the network. Another attempt for a frequency-based attention model was proposed in \cite{yadav2020frequency}. The attention model was adopted from image recognition and could be utilized in any hidden layer of a deep network as the source of query. The attention mechanism uses the latent representations obtained from different layers of the CNN as the information items, thus ruling out the possibility of a localized attention map with respect to the input and individual frequency-bins. Their model also uses a CNN layer to calculate the attention weight. In this approach, and others that similarly employ CNN layers in such way, the information items go through non-linear operations preventing the model from maintaining a one-to-one relation between the attention map and the information items. Therefore as this approach may be successful for some applications, it fails in others where the contribution of each separate information item, in this case individual frequency-bins, is important. Lastly, the notion of using input information items with attention has been used in the area of natural language processing in \cite{wang2016inner}. It should be noted, however, that while the attention mechanism in their work is capable of focusing on individual words, the gradients used for training the attention weights are obtained from the last layers of the model. Adopting a similar mechanism on high dimensional inputs such as spectrograms could result in very complex models which would be hard to train.

Generally, the intuition behind many attention models (in speech-related tasks or otherwise) is to focus on different parts of some latent representation of the input to inform better classification. In these models, the representations are generally learned irrespective of important known information items in the input. Speech depends on frequency content to convey information. In fact, humans have evolved to understand different facts about the source of speech (e.g. identity, intent, emotions, etc.) based on factors such as tone, pitch, and others \cite{hansen_speaker_2015}. The intuition behind our model, which separates it from others, is that by learning to exploit specific frequency-bins in the input that may contain effective task-related information, DNNs can learn to pay more attention to those particular bins to achieve better performance.

\section{Conclusion}
In this paper, a novel attention mechanism is proposed that allows deep learning models to focus on fine-grained information items, namely frequency-bins, without increasing the complexity of the model. Our proposed model, FEFA, uses the spectrogram representations as input and provides a better representation of the spectrogram by attending to each frequency-bin individually. We evaluate our attention mechanism on two tasks of speaker recognition and speech emotion recognition, along with spoken digit recognition. The comparison between models enhanced by FEFA and the original backbone baselines shows consistent improvement in the performance of deep learning models in these tasks. Our analysis shows that the fully connected configuration of FEFA generally outperforms the locally connected configuration. Moreover, we observe that multiple integrations of FEFA does not provide any advantages, and in fact slightly reduces performance, compared to the single integration. We also study the impact of FEFA on training of DNNs and show that not only it has no negative effects on the process, but in some cases it also helps the models to converge faster and stay more stable while doing so. Lastly we test the effect of adding FEFA into the backbone on the robustness of the models. Our experiments show that the models integrated with FEFA exhibit more robustness to different levels and types of synthetic noise. 

\section{Limitations and Future Work}
In the end, to discuss the limitations of our model, we can mention the average pooling step deployed for aggregating frequency information across time. By using such a simple pooling technique, our approach may not be able to gain information such as changes in the values of frequency-bins throughout the duration of the utterance. As a future step we intend to address this weakness by using more complex mechanisms to replace the average pooling for more sophisticated and dynamic aggregation across time \cite{okabe_attentive_2018, xie2019utterance}. Other future steps of this work may include further study of FEFA under more challenging scenarios such as severely imbalanced datasets. Another possible future step of this work would be to exploit FEFA in domain adaptation of DNN models without the need for retraining them. We also intend to study the use of FEFA in creating smaller DNNs by only using the frequency-bins that have the highest weights in the attention map generated by FEFA and removing the other ones. Another interesting possibility is to study the explainability of the attention maps generated by FEFA to better understand its role in performing different audio-related recognition tasks.

\bibliographystyle{IEEEtran}
\bibliography{
              References.bib
              }

\end{document}